\documentstyle[preprint,aps,floats,epsf]{revtex}
\begin{document}
\draft
\title{ Electronic and Structural Properties of C$_{36}$ Molecule
}

\author{Xiaoqing Yu$^{1,2}$, Congjun Wu$^{1,3}$, Chui-lin Wang$^2$, 
and Zhao-Bin Su$^3$ }
\address{
$^1$ Department of Physics, Peking University\\ 
$^2$ China Center of Advanced Science and Technology(World Laboratory),\\
P.O. Box 8730, Beijing 100080, China\\
$^3$ Institute of Theoretical Physics, Chinese Academy of Sciences,\\
P.O. Box 2735, Beijing 100080, China
}
\maketitle

\begin{abstract}
The extended SSH model and Bogoliubov-de Gennes(BdeG) formalism are applied
to investigate the electronic properties and stable lattice configurations
of C$_{36}$.
We focus the problem on the molecule's unusual $D_{6h}$ symmetry.
The electronic part of the  Hamiltonian without Coulomb interaction
is solved analytically.
We find that the gap between HOMO and LUMO is small
due to the long distance hopping between the 2nd and 5th layers.
The charge densities of HOMO and LUMO are mainly distributed in the two layers,
that causes a large splitting between the spin triplet and singlet excitons.
The differences of bond lengths, angles and charge densities
among the molecule and polarons are discussed.
\end{abstract}

\vspace*{.8cm}

{\bf Contact Author}: Xiaoqing Yu

{\bf E-mail}: xyu4@students.uiuc.edu

\newpage

\section{INTRODUCTION}

\noindent
Recently, a new member of fullerenes, C$_{36}$, was synthesized by 
arc-discharge method and purified in bulk quantities\cite{Exp}.
Up to now, it is the smallest fullerene ever discovered.
The C$_{36}$ molecule is more curved than C$_{60}$ because of
the adjacent pentagonal rings and the small number of carbon atoms\cite{Exp}.
This feature suggests stronger electron-phonon interaction and
possible higher superconducting transition temperatures
in the alkali-doped C$_{36}$ solids than those of C$_{60}$\cite{ElPh}.
The Solid-State Nuclear Magnetic Resonant experiment suggested that the
most favorable configuration of C$_{36}$ molecule has $D_{6h}$ symmetry
\cite{Exp}.
This confirmed the results of the early {\it ab initio} 
pseudopotential density functional calculations\cite{Chem}, which indicated
that the $D_{6h}$ structure is 
one  of the most energetically favorable structures among several possibilities.

Looked in a direction perpendicular to the six-fold 
axis(Fig.1), the molecule is composed of six parallel layers.
On each layer, six carbon atoms lie at the vertexes of a hexagon.
The edges of the hexagons of the 1st, 2nd, 5th, 6th layers are parallel,
while those of the two hexagons in the middle two layers 
are turned by an angle of $30^\circ$.
In addition to  this special structure, the six-fold principal axis
is distinct form the ordinary five-fold principal axis
in the cases of C$_{60}$ and C$_{70}$ and little attention has been paid to it.
In this paper, we discuss both the electronic and structural properties
of C$_{36}$ molecule, emphasizing the interesting properties brought 
by this unusual $D_{6h}$ symmetry.

\begin{figure}
\epsfxsize=6cm
\centerline{\epsffile{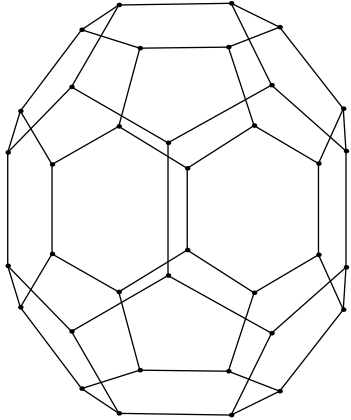}}
\caption{
The geometrical figure of C$_{36}$ molecule.
}
\end{figure}
We employ a simple and elegant model , extended SSH model,
which is successful in treating C$_{60}$ and C$_{70}$\cite{Wang,Tian}. 
The BdeG formalism is performed 
to obtain the stable lattice configuration
and the corresponding electronic states self-consistently.
The electronic part of Hamiltonian without Coulomb interaction 
can be solved analytically by methods of group theory.
We find that the electron densities of HOMO($B_{1u}$) and LUMO($B_{2g}$) 
are zero at the two middle layers.
Furthermore, the HOMO and LUMO are mainly confined in the 2nd and 5th layers
with a possibility of $90\%$.
The gap between them is 
considerably small, compared to that of C$_{60}$ and C$_{70}$, 
because the splitting of HOMO and LUMO is due to a long-distance 
hopping in our case.
We find the large splitting of the triplet and singlet excitons due to
more localized HOMO an LUMO in comparison with C$_{60}$ and C$_{70}$.
As a result, the triplet exciton's energy is very small and possible experiments
are suggested to test this phenomenum.

The geometrical figures of C$_{36}$ molecule and polarons are discussed and
bond lengths and angles are calculated.
The charge density on each layer are given. 
We find that the differences of charge densities among polarons
and molecule are mainly in the 2nd and 5th layers.

The following sections are arranged as: in Sec.$\rm I\!I$, 
the extended SSH model is introduced; 
in Sec.$\rm I\!I\!I$, 
we analytically solve the electronic  part of the Hamiltonian without
Coulomb interaction;
in Sec.$\rm I\!V$,
the whole Hamiltonian is solved self-consistently; 
in Sec.$\rm V$, results and discussions are presented;
and conclusions are made in the final section.

\section {MODEL}
\begin{figure}
\epsfxsize=6cm
\centerline{\epsffile{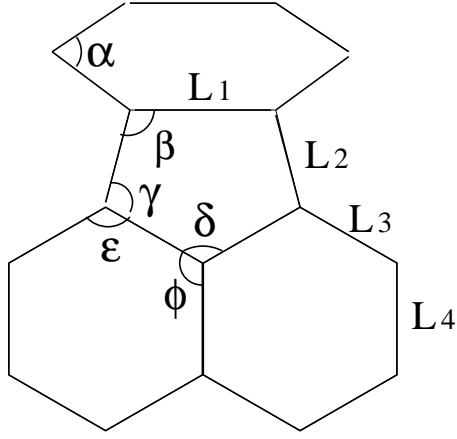}}
\caption{
The geometrical figure of C$_{36}$ molecule.
}
\end{figure}

\noindent
The Hamiltonian of the extended SSH model to the C$_{36}$ molecule 
is written as:
\begin{equation}
      H = H_0 +H_{int}+H_{elas} 
\end{equation}
The first term $H_0$ of Eq.(1) is the hopping term of $\pi$-electrons.
\begin{eqnarray}      
H_0 = &-&\sum_{\langle i,j \rangle} [t_0-\alpha(l_{ij}-l_0)](c_{i\sigma}^\dagger
c_{j\sigma}+h.c.)
-{\sum_{\langle i,j \rangle}}^\prime t_1(c_{i\sigma}^\dagger c_{j\sigma}+h.c.)
\nonumber\\
&-&{\sum_{\langle i,j \rangle}}^{\prime\prime} t_2(c_{i\sigma}^\dagger 
c_{j\sigma}+h.c.)
\end{eqnarray}
where the three terms describe hopping between the nearest neighbors,
the next nearest neighbors and the third neighbors respectively,
with the corresponding hopping integrals $t_0,t_1$ and $t_2$. 
The influence of electron-phonon coupling is only included 
in the first term of $H_0$, since those of the last two terms are 
much smaller than $t_0$. 
The third term is important here because it changes 
the accidental degeneracy of  HOMO and LUMO states ,
which will be  explained in the next section. 
The larger distance terms are neglected, 
as they are small and do not bring any new effects.

The second term of Eq.(1) is the screened Coulomb interaction expressed
in the Hubbard model. 
\begin{equation}     
      H_{int} = U {\sum_i n_{i\uparrow}n_{i\downarrow}}+
       V {\sum_{\langle i,j \rangle,\sigma,\sigma^\prime}}
         n_{i\sigma}n_{j\sigma^\prime}    
\end{equation}
where $U$ is the strength of the on-site interaction, and $V$ is 
that between the nearest neighbors.

The third term of Eq.(1) is the elastic energy of the lattice. This
term is composed of three parts,
\begin{equation}
H_{elas}=\frac{1}{2}K_1{\sum_{\langle i,j \rangle}}(l_{ij}-l_0)^2
+\frac{1}{2}K_2\sum_i {d\theta_{i5}}^2
+\frac{1}{2}K_3\sum_i {d\theta_{i6}}^2,
\end{equation}
where the first part describes the spring energy of bond-length terms,
and the next two terms describe the spring energy of the angular terms.
$K_i(i=1\sim 3)$ are the elastic constants for these different kinds of 
lattice vibrations. $d\theta_{i5},d\theta_{i6}$ are bond angle 
deviations from the original angle $108^\circ,120^\circ$. e.g. $d\theta
_{i5}=\theta_{i5}-108^\circ$. The first summation is taken up over all
nearest pairs. The second is of all interior angles of pentagons. And the 
third is of all interior angles of hexagons.

Since there are no enough experiment data to determine the 
semiempirical parameters, 
we set the values in the scope of fullerenes such as C$_{60}$ and C$_{70}$
which can produce reasonable results. 
We adjust them to give the gap between HOMO and LUMO
and the bond length consistent with those of
the pseudopotential density functional approach.
We take $t_0=2.5$ eV, $\alpha=5.6$ eV/\AA, $L_0=1.55$ \AA, 
$K_1=47$ eV/ \AA$^2$, $K_2=8$ eV/rad$^2$, $K_3=7$ eV/rad$^2$, $t_1=0.168 t_0$.
In fact, the particular values, except for the value of $t_2$, would not
affect the physics too much. 
Based on the calculation of the third-nearest $\pi$-orbit integral,
$t_2$ was estimated to be 0.111$t_0$.

\section{ ANALYTICAL RESULT OF THE Electronic HAMILTONIAN
without Coulomb Interaction}
\small
\begin{table}
\caption{The analytical solution of $H_0$ without $t_1$ and $t_2$ terms.
} \label{tab:analy}
\begin{center}
\begin{tabular}{rrccc}   
 m,p~~~~~~~   &E(eV)   &$c_{mp,1}$     &$c_{mp,2}$    &$c_{mp,3}$   \\  \hline       
 $0,+$$(A_{1g})$     &-9.51  &0.581 &0.574  &0.578      \\ 
                &-5.74  &0.791 &-0.232 &-0.566      \\ 
                &5.57   &0.190 &-0.786 & 0.589      \\
 $0,-$$(A_{2u})$   &-8.43  &0.816 &0.504 &0.284\\ 
                &-3.53  &-0.565 &0.589 &0.578 \\ 
   	        &8.43   &-0.124 &0.632 &-0.765\\
$\pm$1,+ $(E_{1g})$ &-6.28  &0.647 &0.654  &0.340 $\pm$ 0.196i \\ 
                   &-1.60  &0.744 &-0.428 &-0.445 $\mp$ 0.257i \\       
                   &7.66   &0.145 $\mp$ 0.084 i&-0.540 $\pm$ 0.312 i&0.763\\ 
$\pm1,-$ $(E_{1u})$ &-8.09 &0.329$\mp$0.190 i & 0.534$\mp$0.308 i &0.690 \\ 
                 &-3.25 &0.882 &-0.015 &-0.409$\mp$0.236 i\\ 
                 &4.97  &-0.281 &0.787& -0.476$\mp$ 0.275 i\\ 
$\pm2,+$ $(E_{2g})$     &-5.44 &0.096$\mp$0.169 i &0.289$\mp$0.501i &0.792\\ 
                   &0.269  &0.573 &0.589 &-0.285$\mp$0.494 i \\ 
                   &5.40 &0.796 &-0.565 &0.108$\pm$0.187 i \\ 
$\pm2,-$ $(E_{2u})$   &-3.07 &0.379 &0.819 &0.216$\pm$0.374 i \\ 
                   &3.20 &0.727 &0.026 &-0.343$\mp$0.595 i \\ 
                   &6.25 &0.287$\mp$0.496 i &-0.287$\pm$ 0.497 i & 0.585\\ 
$3,+$ $(B_{2g})$      &-1.13 &0.357 &0.934 &0       \\ 
   $(B_{1g})$      &3.08  &0   &0   &1         \\
   $(B_{2g})$      &7.73  &0.934  &-0.357  &0 \\ 
$3,-$ $(B_{2u})$      &-3.08 &0  &0    &1 \\ 
   $(B_{1u})$      &-1.13 &0.357 &0.934 &0\\ 
   $(B_{1u})$      &7.73  &0.934 &-0.357 &0 \\    
\end{tabular}
\end{center}
\end{table}

Fully exploiting the high $D_{6h}$ symmetry, we solve $H_0$ algebraically
as in the case of C$_{60}$\cite{Exac,Yang}. 
For simplicity, we temporarily ignore $t_1$ and $t_2$ terms and 
will discuss their effects in detail later. 
The thirty-six $\pi$ orbits form a 36$\times$36 representation
of the $D_{6h}$ group.
It can be reduced into the sum of following irreducible representations:
\begin{equation}
\{3A_{1g}\oplus B_{1g}\oplus 2B_{2g}\oplus 3E_{1g}\oplus 3E_{2g}\}
\oplus\{2B_{1u}\oplus 3A_{2u}\oplus B_{2u}\oplus 3E_{1u}\oplus 3E_{2u}\}
\end{equation}
The even-order axis C$_6$ brings properties different  
from those odd-order axis $C_5$ characterized in C$_{60}$ and C$_{70}$, 
in the way that it has 1-D representations of kind B.

We reduce this problem by $D_{6h}$'s subgroup $C_{6h}$, using the 
quantum number m(=0,$\pm1$,$\pm2$,3) which corresponds to the irreducible 
representation of $C_{6h}$, and $P(=\pm1)$ to parity. The thirty-six $\pi$ 
orbits are recombined as:
\begin{equation}
|\Psi_{mp}^{(l)}\rangle=\sum_i \eta^{mi}\{|l,i\rangle+P|7-l,i\rangle\}.
\end{equation}
where $|l,i\rangle$ represents the $i$th($i=1\sim 6$) carbon atom's $\pi$ orbit
in the $l$th layer. 
Here $l$ is only ranged for 1 to 3, and  $\eta=e^{i\pi/3}$

The energy eigenstate wavefunction $\Phi_{mp}^l$ can be expanded with
these base vectors.
\begin{equation}
\Phi_{mp}^i=\sum_{l=1,3} c^i_{mp,l} \Psi_{mp}^l
\end{equation}
Consequently, $H_0$ can also be reduced to 
3$\times$3 matrices in the subspace which is spanned by the new bases:
\begin{equation}
H_{mp}=\left[
\begin{array}{ccc}
-t_a(\eta^{-m}+\eta^m) & -t_b & 0 \\
-t_b & 0 & -t_c(1+\eta^{-m}) \\
0 & -t_c(1+\eta^m) & -t_dP\eta^{3m}
\end{array}  \right]
\end{equation}
Because of the $D_{6h}$ symmetry, there are only four different kinds
of bond lengths ($L_a\sim L_d$), see Fig.2. The corresponding nearest
hopping integrals are $t_i=t_0-\alpha(L_i-L_0) (i=a\sim d)$.
The energy eigenvalues $E^l_{pm}$ are determined by
\begin{eqnarray}
\lambda^3+A\lambda^2+B\lambda+C=0 \nonumber
\end{eqnarray}
where
\begin{eqnarray}
&& A=2 t_a \cos {m\pi \over 3} +(-1)^m p t-d, \nonumber \\
&& B= 2 \cos{m\pi \over 3} [ (-1)^mpt_a t_d -t_c^2] -(2t_c^2+t_b^2),\nonumber \\
&&
C=-4 \cos{m\pi \over 3}(1+\cos{m\pi \over 3})t_a t_c^2 -(-1)^m p t_b^2 t_d
\nonumber
\end{eqnarray}
Equation(9) can be solved analytically, so does the eigenvectors.
Here we would not list the complicated analytical results.
Instead, we present numerical results by using
parameters sets,$t_a=3.30$eV, $t_b=2.95$eV, $t_c=3.24$eV, 
$t_d=3.08$eV,
which are determined by the bond lengths $L_i$.
The coefficients $c_{mp,l}^i$ of the energy eigenstate wavefunction $\Phi_{mp}^i$
and their irreducible representations of $D_{6h}$ that they belong to 
are showed in Table 1.

\normalsize
In the C$_{36}$ molecule, because the hexagons in two middle layers
turn an angle of $30^\circ$ to the four hexagons at two ends, 
the wavefunctions of $B_{2u}$ and $B_{1g}$ energy eigenstates are 
entirely composed of the atom orbits on the two middle layers, 
while those of $B_{1u}$ or $B_{2g}$ energy eigenstates 
only have amplitude on the other four layers.
There are six energy eigenstates of the B kind representations.
The 13th($B_{2u})$, 22nd($B_{1g})$, 
the  accidentally degenerate 35th($B_{1u}$) and 36th($B_{2g}$) levels
lie either far below or above the fermi level,
so they are of no interest.
There are also another pair of accidentally degenerate 18th($B_{1u}$) and 
19th($B_{2g}$) levels which are half filled.
We also notice in Table 1. that the 18th($B_{1u}$) and 19th($B_{2g}$)
electrons are distributed in the 2nd and 5th layers with $90\%$ possibility
and $10\%$ in the 1st and 6th layers.

When $t_1(16.8\%t_0)$ and $t_2(11.1\%t_0)$ are considered, $H_{mp}$ have 
a different form.
The long distance hopping($t_2$) coupling the 2nd and 5th layers results
in $B_{1u}$ and $B_{2g}$'s splitting.
This is the reason that we introduce much longer distance
hopping term than usual.
Because the charge densities of $B_{1u}$ and $B_{2g}$ are mainly
distributed in the  2nd and 5th layers, 
the gap between HOMO($B_{1u}$) and LUMO($B_{2g}$) is approximately $2t_2$. 
When the electron-electron interaction is taken into account, 
the  electronic Hamiltonian
has to be solved in the HF mean field theory with self-consistent method. 
The energy levels of ground state configuration are shown in Fig.3. 
We can see that in Fig.3 and Table 1,
the relative positions of the energy levels only change slightly,
except the splitting of $B_{1u}$ and $B_{2g}$.
In Fig.2, the separation between the 16th, 17th levels($E_{1g}$)
and HOMO is about 0.2eV and that between the 20th, 21st levels
($E_{2g}$) and LUMO is 1.3eV.

\section {NUMERICAL RESULT OF THE MOLECULE}
Under the adiabatic approximation,
we apply the BdeG formalism to the Hamiltonian of the molecule to obtain 
the stable lattice configuration and the corresponding electronic structure.
Molecular dynamical procedure is used to gradually approach
the minimum of the potential surface from an initial lattice configuration.
The following two equations are used, 
\begin{eqnarray}
F_{i\sigma} & = & -\frac{d~V_{eff}}{d~x_{i\sigma}} ~~~~~~~~~~~~~
~~~~ (\sigma =1-3) \\
v_{i\sigma} & = & \frac{d~x_{i\sigma}}{d~t}
\end{eqnarray}
where $V_{eff}$ is the effective lattice potential that includes the 
elastic energy of the lattice and the electron-lattice interaction 
energy, both of which depends on the lattice coordinates. $F_{i\sigma}$
is the $\sigma$ component of the effective force acted on the $ith$ atom under 
this potential. $x_{i\sigma}$ is the $\sigma$ coordinate of the $ith$ atom.

In each step of the dynamical procedure, we solve the electrical part 
of the Hamiltonian under the lattice configuration given by the last step.
The HF mean field theory is performed to decouple the electron-electron 
interaction. Then the electronic states and wavefunctions are obtained 
self-consistently.
\begin{eqnarray}
H_{MF}^{el} & = & H_0+U\sum_i \{ \langle n_{i\uparrow}\rangle n_{i\downarrow}+
\langle n_{i\downarrow}\rangle n_{i\uparrow}-
\langle n_{i\uparrow}\rangle\langle n_{i\downarrow}\rangle\}
\nonumber \\
 & & 
+V\sum_{\langle{i,j}\rangle,\sigma,\sigma^\prime}
\{ \langle{n_{i\sigma}}\rangle n_{j\sigma^\prime}+
 \langle{n_{j\sigma^\prime}}\rangle n_{i\sigma}-
 \langle{n_{i\sigma}}\rangle\langle{n_{j\sigma^\prime}}\rangle\}
 \nonumber \\
 & &
-V\sum_{\langle i,j \rangle,\sigma}
\{ \langle C^\dagger_{i\sigma}C_{j\sigma}\rangle 
C^\dagger_{j\sigma} C_{i\sigma}+
\langle C^\dagger_{j\sigma} C_{i\sigma} \rangle
C^\dagger_{i\sigma}C_{j\sigma} -
\langle C^\dagger_{i\sigma}C_{j\sigma}\rangle
\langle C^\dagger_{j\sigma} C_{i\sigma} \rangle\}
\end{eqnarray}
The ground state and the lowest triplet and singlet excitons are investigated.
There are four lowest exciton configurations $A,B,C$ and $D$, see Fig. 4.
Due to Coulomb interaction, $C$ and $D$ are mixed to give singlet and triplet,
\begin{eqnarray}
\langle C|H_{int}&&|D\rangle = 
-\sum_i U \psi^*_{\alpha\uparrow}(i) \psi^*_{\beta\downarrow}(i)
         \psi_{\beta\uparrow}(i)  \psi_{\alpha\downarrow}(i)\nonumber \\
&&- V\sum_{\langle ij \rangle} \{ \psi^*_{\alpha\uparrow}(i) 
         \psi^*_{\beta\downarrow}(j)
             \psi_{\beta\uparrow}(i) \psi_{\alpha\downarrow}(j)
+               \psi^*_{\alpha\uparrow}(j) \psi^*_{\beta\downarrow}(i)
             \psi_{\beta\uparrow}(j) \psi_{\alpha\downarrow}(i) \}                         
\end{eqnarray}
where $\alpha$ and $\beta$ denote the 18th and 19th energy level respectively.

\begin{figure}
\epsfxsize=15cm
\centerline{\epsffile{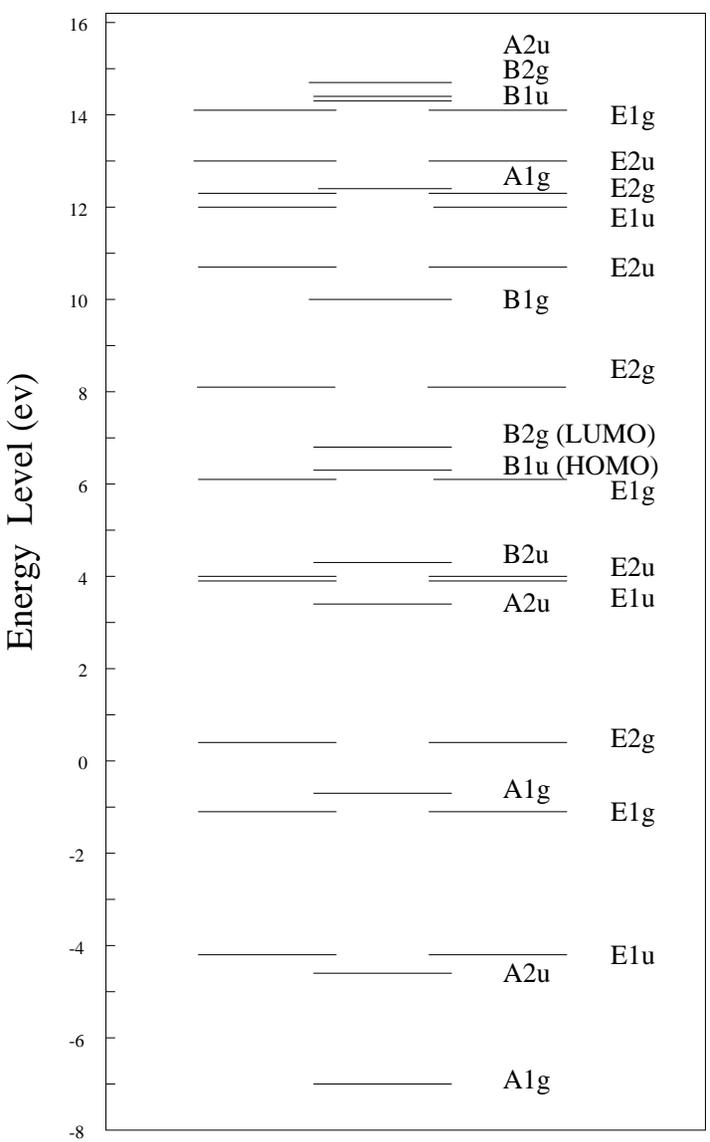}}
\caption{
Energy levels o the ground state C$_{36}$ molecule.
}
\end{figure}
 
\begin{figure}
\epsfxsize=15cm
\centerline{\epsffile{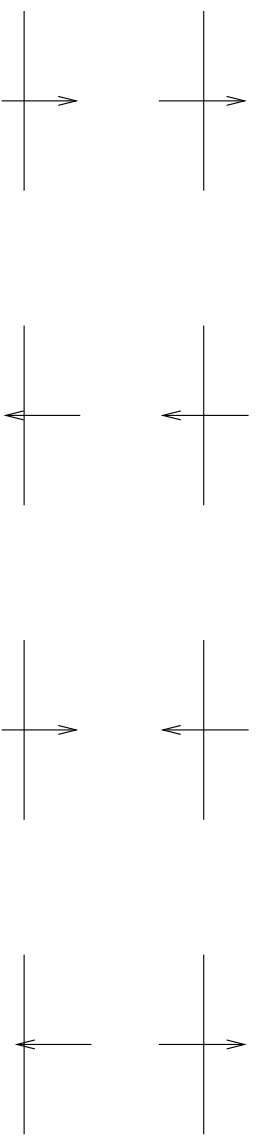}}
\caption{
The four possible configurations of the lowest excitons.
}
\end{figure}

\section { RESULTS AND DISCUSSION}
\begin{table}[h]
\begin{center}
\begin{tabular}{crrrrrr}   
              & 1     &2     &3     &4    &5    &6       \\ \hline       
C$_{36}^{2-}$ &1.053  &1.131 &0.991  &0.991 &1.131 &1.053 \\ 
C$_{36}^{-}$  &1.042  &1.057 &0.985  &0.985 &1.057 &1.042 \\ 
C$_{36}$      &1.031  &0.981 &0.988  &0.988 &0.981 &1.031 \\ 
C$_{36}^+$    &1.021  &0.902 &0.992  &0.992 &0.902 &1.021 \\ 
C$_{36}^{2+}$ &1.013  &0.823 &0.997  &0.997 &0.823 &1.013 
\end{tabular}
\end{center}
\caption{The ground state charge density per site on each layer of
ions and molecule.}
\end{table}

\begin{table}[h]
\begin{center}
\begin{tabular}{cllll}   
      & $L_1$ &$L_2$ &$L_3$ &$L_4$ \\ \hline         
SSH(C$_{36}^{2-}$)&1.414   & 1.453  & 1.418   & 1.447   \\ 
SSH  (C$_{36}^-$)& 1.410   & 1.462  & 1.418   & 1.447   \\ 
SSH(C$_{36}$) & 1.407   & 1.470  & 1.418   & 1.447   \\ 
SSH(C$_{36}^+$)   &1.404   & 1.477  & 1.417   & 1.447   \\ 
SSH(C$_{36}^{2+}$)&1.402   & 1.484  & 1.417   & 1.448   \\ 
LDA(C$_{36}$) & 1.41    & 1.48   & 1.43    & 1.43   
\end{tabular}
\end{center}
\caption{The bond length(\AA) of ions and molecule.}
\end{table}

\begin{table}[h]
\begin{center}
\begin{tabular}{ccccccc}  
   & $\alpha$ &$\beta$ &$\gamma$ &$\delta$ & $\varepsilon$ & $\phi$ \\ \hline         
C$_{36}$ &$120.00^\circ$ &$107.64^\circ$ & $108.20^\circ$  & $108.29^\circ$  
    & $119.42^\circ$ & $119.24^\circ$ 
\end{tabular}
\end{center}
\caption{The bond angle within SSH model.}
\end{table}  

\noindent
The configurations and energy of ground and low excited states are given
below,
\begin{mathletters}
\begin{eqnarray}
E_1^{el}&=&-64.25~eV ~~~~~~~~~~~~~ |\Phi_1\rangle
~~\mbox{(spin singlet ground state)}, \\
E_2^{el}&=&-64.00~eV ~~~~~~~~~~~~~ |\Phi_2^{1\sim3}\rangle 
=|A\rangle, |B\rangle, {1\over {\sqrt 2} } (|C\rangle+|D\rangle)
~~\mbox  {(spin triplets)}, \\
E_3^{el}&=&-63.37eV, ~~~~~~~~~~~~~|\Phi_3\rangle = {1\over {\sqrt 2} }
(|C\rangle-|D\rangle) ~~\mbox{(spin singlet)}.
\end{eqnarray}
\end{mathletters}
We can see that the ground state $\Psi_1$ is a spin singlet rather than triplet
after the long distance hopping $t_2$ and electron-electron interaction are
taken into account,
which is an improvement of the prediction of simple H\"{u}ckel theory,
and there is no need to add the  hybridizing of $\sigma$ and $\pi$ bonds
to investigate the qualitative physics of C$_{36}$ as discussed in Ref.\cite{C36}.

The lowest excited states $\Psi_2^{1-3}$ are spin triplet excitons,
which are only about $0.25$eV above the ground states.
The energy of singlet exciton $\Psi_3$ is very high, 
and the splitting between the triplet and singlet excitons is about 0.63eV,
which is much larger than 0.2eV in the cases of C$_{60}$ and C$_{70}$.
This is because the electrons of HOMO and LUMO are more localized,
$90\%$ in the 2nd and 5th layers and $10\%$ in the 1st and 6th layers. 
Compared to the C$_{36}$ case,
the electron densities of HOMO and LUMO in C$_{60}$ and C$_{70}$
are distributed more uniformly over all sites,
so the splitting of the singlet and triplet exciton due to the Coulomb 
interaction is relatively small.
The fact that the triplet exciton has low energy can be verified through
experiment observation:
When illuminated by external light, the C$_{36}$ can be excited to
the state of singlet exciton and then may transit to triplet exciton 
through nonradiative decay.
The transition rate to ground state is slow because of
the different spin configuration and small energy splitting.
We predict that the triplet excitons are metastable states,
and the phosphorescence  phenomena is possible to be observed.
The triplet exciton is paramagnetic 
while the ground state is a diamagnetic singlet.
So the magnetic susceptibility is increased upon illumination 
of the external light.
Electron Spin Resonance(ESR) experiment can be performed to detect the
triplet exciton states.
When ESR is performed to a solution sample,
the usual unimportant magnetic dipole interaction between
electrons becomes crucial because its anisotropy can smear the resonance peak.
However, this interaction is decayed rapidly with distance as $R^{-3}$.
The electrons in HOMO and LUMO are mainly uniformly distributed on
12 carbon atoms in two layers,
the possibility of their short distance is considerably small compared to 
other small organic molecule, such as naphthalene.
So the resonance peak is possible to be observed.  
The phone absorption and luminescence spectra experiments can
be perform to test the lowest singlet exciton's energy.

The shape of the C$_{36}$ molecule is an ellipsoid with high aspect ratios.
Our calculation shows that the distance between the 1st and 6th layers
is $5.2A^\circ$,
and that between the parallel vertical hexagon planes is $4.2A^\circ$.
The charge density, bond lengths and bond angles of C$_{36}$ molecule 
and ions are calculated and shown in Table 2, 3 and 4.
We note that because of the $D_{6h}$ symmetry of the molecule,
the 1st and 6th layers are regular hexagons, 
with each internal angle $120^\circ$ and the shortest bonds 
($L_1$) $1.41A^\circ$. 
The six vertical hexagons around the equator 
are slightly deformed due to the drag of the 1st and 
6th layers from each side.
$L_4$ is $2.8\%$ longer than $L_1$ and $2\%$ than $L_3$.
The $L_2$ bonds are relatively weaker than those in  hexagons, because they
couple  the hexagons in the two ends with the hexagons around the equator.
$L_2$ is $4.4\%$ longer than $L_1$.
This result is consistent with that of LDA except that the $L_4$ is as long
as $L_3$ in LDA,
and we think that our result is more reasonable.

The bond lengths of $L_3$ and $L_4$ of ions differ from those of molecule
very slightly.
But the negative polarons' $L_1$ is longer and positive polaron' is shorter
than the neutral molecule's, while the $L_2$ changes in the opposite way.

The reason is that the wavefunctions of the electron in $B_{1u}$ and $B_{2g}$ 
has opposite sign between two nearest sites connected by $L_1$, same sign 
by $L_2$, and have zero amplitude on the 3rd, 4th layers. 
According to the contribution f these two states form hopping term,
the sites connected by $L_1$ are repulsive, but those connected by $L_2$ 
are attractive. 
The angles in the pentagons and hexagons around the equator 
deviate from $108^\circ$
and $120^\circ$ slightly, 
which justifies the validity of $H_{elas}$.
The bond angles in polarons differ from those of molecule very slightly,
so we omit them in Table 4.
Form C$^{2+}_{36}$ polaron to C$^{2-}_{36}$, the adding electrons
distribute mostly in the 2nd and 5th layers.
This is because of the  particular charge density distribution of the
18th and 19t levels.

\section {CONCLUSION}
In this paper, we have carefully studied the effect of $D_{6h}$ symmetry on 
C$_{36}$'s electronic properties under the extended SSH model. 
A small gap between HOMO($B_{2u}$) and LUMO($B_{1g}$) is obtained due to 
long distance hopping. 
The large splitting of the spin triplet and singlet lowest excitons,
the differences of bond lengthes and electron density between molecule 
and polarons are discussed as results brought by the more localized HOMO 
and LUMO and their special symmetries.
Possilbe experiments are suggested.

\bigskip
\centerline {\bf  ACKNOWLEDGMENT}

We thank Dr. Xi Dai for his helpful discussions.

\end{document}